\documentclass[aps,prd,groupedaddress]{revtex4}
\usepackage{graphicx,epsf,amssymb,amsmath,latexsym}
\usepackage{epsfig}
\def\ben{\begin{equation}}
\def\een{\end{equation}}
\def\bena{\begin{eqnarray}}
\def\eena{\end{eqnarray}}

\begin{document}

\title{Strong Interaction Dynamics from Spontaneous Symmetry Breaking of Scale Invariance}

\vspace{.3in}

\author{E.I. Guendelman}
\email{guendel@bgu.ac.il}

\affiliation{Physics Department, Ben-Gurion University of the Negev, Beer-Sheva 84105, Israel }

\vskip.3in
\begin{abstract}
%

Using the mechanism of spontaneous symmetry breaking of scale
invariance obtained from the dynamics of maximal rank field
strengths, it is possible to spontaneously generate confining
behavior. Introducing a dilaton field, the study of non trivial
confining and de-confining transitions appears possible. This is
manifest in two ways at least: One can consider bags which contain
an unconfined phase in the internal region  and a confined phase
outside and also one obtains a simple model for deconfinement at
high Temperature from the finite temperature dynamics of the
dilaton field.
\end{abstract}

\maketitle
%

\setcounter{equation}{0}
\section{Introduction}

Strong interaction dynamics involves many remarkable phenomena:
first, QCD at very short distances has the asymptotically free
property \cite{free}. In contrast to this, in the infrared region,
we are should obtain the confined phase, where only color singlets
survive, since no free quarks or gluons have been observed so far.
Also at high temperatures, we expect a deconfinement phase
transition \cite{deconf}.

Although the asymptotically free property is clearly understood
theoretically, the  confinement is not. Lattice gauge theory
provides nevertheless numerical evidence for  confinement in the
context of QCD \cite{lattice}.

Given the theoretical difficulties concerning the description of
the confining phase, several phenomenological approaches have been
developed , for example the Cornell potential for  heavy quarks
\cite{Cornell} and the MIT bag model \cite{MIT}, which gives a
comprehensive description of hadron spectroscopy.

In Ref. \cite{Guendelman}  I proposed a model where the
spontaneous symmetry breaking (ssb) of scale invariance induces an
effective dynamics which gives rise to confinement. Here the ssb
of scale invariance originates from the dynamics of  maximal rank
gauge fields. The resulting theory that results in fact satisfies
the requirements studied by 't Hooft \cite{pert-conf} for
perturbative confinement .

More explicit computations in the context of this model
\cite{Guendelman-Gaete} have shown that it gives rise to the
Cornell confining potential for static sources\cite{Cornell}. This
model can be used also to study the interplay between confinement
effects and ssb of gauge symmetry \cite{interplay} .

In this paper we will see that a simple generalization of
\cite{Guendelman} through the introduction of a dilaton field
allows to enlarge the set of phenomena described. In addition to
confinement we can also obtain bag structures where a deconfined
phase is obtained in a central region, while confinement is
obtained outside. Also , a simple description of the deconfinement
phase transition at high temperature is obtained.

\section{Confinement from Spontaneous Symmetry Breaking of Scale Invariance}

In this section, we discuss the connection between the scale
symmetry breaking and confinement, introduced in
Ref.\cite{Guendelman}. For this purpose we restrict our attention
to the action
\begin{equation}
S_{YM} = \int {d^4 } x\left( - \frac{1}{4}F_{\mu \nu }^{a} F^{a\mu
\nu}
 \right)\ ,\label{YM}
\end{equation}
where $ F_{\mu \nu }^a  = \partial _\mu  A_\nu ^a - \partial _\nu
A_\mu ^a + e f^{abc} A_\mu ^b A_\nu ^c$, where $e$ is the gauge
coupling constant.
 This theory is invariant
under the scale symmetry
\begin{equation}
A^{a}_\mu  \left( x \right) \mapsto A_\mu ^ {a\prime}  \left( x
\right) = \lambda^{-1}\, A^{a}_\mu  \left( \lambda\, x \,\right),
\end{equation}
where $\lambda$ is a constant.

Let us introduce  the following action, which is at this point
equivalent to the action above
\begin{eqnarray}
{\mathcal{S}_1}\left[\, \omega\ ,  A^{a}_\mu\ \right]
 =&& \int {d^4 } x\left[\,  - \frac{1}{4}\omega ^2
+\frac{1}{2}\omega \sqrt { - F_{\mu \nu }^{a}\, F^{a\mu \nu } }
\right]\,  \label{ma}
\end{eqnarray}
where, $\omega$  is an
auxiliary scalar field,  its scaling transformation is
\begin{eqnarray}
&&\omega  \mapsto \lambda ^{-2} \omega \left( \lambda\, x
\,\right)\label{cr25c}
\end{eqnarray}

Upon solving the equation of motion associated to the variation
with respect to $\omega$, we obtain that $\omega = \sqrt { -
F_{\mu \nu }^{a}\, F^{a\mu \nu }}$ and inserting back into (3), we
obtain once again (1).

Let us now declare that $\omega$ is not a fundamental field, but
that instead
\begin{equation}
\omega = \epsilon^{\lambda\mu\nu\rho}
\partial_{[\,\lambda}\, A_{\mu\nu\rho \,]}
\end{equation}
Now the variation with respect to $ A_{\mu\nu\rho }$ gives rise to
the equation:
\begin{eqnarray}
 \epsilon^{\lambda\mu\nu\rho} \partial_{\lambda}
 \left[\ \omega - \sqrt { - F_{\mu \nu }^{a}\, F^{a\mu \nu } }\right]\ = 0
\end{eqnarray}
which is satisfied by
\begin{eqnarray}
\omega = \sqrt { - F_{\mu \nu }^{a}\, F^{a\mu \nu }} + M
\end{eqnarray}
Here $M$ is a space time constant which produces the ssb of scale
invariance and it is in fact associated with the spontaneous
generation of confining behavior. This can be seen by considering
the equations of motion in the case $M\neq 0$
\begin{equation}
 \nabla _ \mu
\left[ {\left( {\sqrt { - F_{\alpha \beta }^a F^{a\alpha \beta } }
+ M} \right)\frac{{F^{a\mu \nu } }}{{\sqrt { - F_{\alpha \beta }^b
F^{b\alpha \beta } } }}} \right] = 0. \label{cr50}
\end{equation}
Then, assuming spherical symmetry and time independence, we obtain
that, in addition to a Coulomb like piece, a linear term
proportional to $M$ is obtained for $A^{a 0}$. A quantum
computation also confirms this result \cite{Guendelman-Gaete}.

Furthermore, these equations are consistent with the 't Hooft
criteria for perturbative confinement. In fact in the infrared
region the above equation implies that
\begin{equation}
{F^{a\mu \nu } = - M \frac{{F^{a\mu \nu } }}{{\sqrt { - F_{\alpha
\beta }^b F^{b\alpha \beta } } }}}. \label{cr51}
\end{equation}
plus negligible terms in the infrared sector. Interestingly
enough, for a static  source, this automatically implies that the
chromoelectric field has a fixed amplitude. Confinement is obvious
then, since in the presence of two external oppositely charged
sources, by symmetry arguments, one can see that such a constant
amplitude chromoelectric field must be in the direction of the
line joining the two charges. The potential that gives rise to
this kind of field configuration is of course a linear potential.
Notice that the above equation implies that $M<0$, otherwise the
electric field would be antiparallel to itself. The string tension
between static quarks of opposite charge is proportional to the
absolute value of $M$.

A configuration with a net charge is seen from the above equation
to produce, at very large distances from the source, an electric
field which is radial but constant in strength. This will in fact
produce an infrared divergent total energy for the system. In
general any non color singlet will be un-physical because these
systems will have infinite energy. Furthermore such a configuration
with net charge will detect and be attracted to any opposite
charge no matter how far this will be by a linear potential, so a
system with net charge can never be isolated.

\section{Generalized Models Allowing Bag Structures and Deconfinement Phase Transitions}

We now consider a model that generalizes that of the previous
section and which includes also a scalar field $\phi$
\begin{eqnarray}
{\mathcal{S}_G}\left[\, \delta\ ,\phi\ ,  A^{a}_\mu\ ,\omega
\,\right]
 = \int {d^4 } x\left[\,  - \frac{1}{4}\delta^{2}
+\frac{1}{2}\delta \sqrt { - F_{\mu \nu }^{a}\, F^{a\mu \nu } }
\right] +\int {d^4 } x \left[ \phi^{2}\,\left(\, \omega
-\delta\,\right)\,+\frac{1}{2}\partial_{\mu}\phi
\partial^{\mu}\phi - V(\phi)
-\frac{1}{2}\omega^{2}\right]
\end{eqnarray}
Now  $\delta$ is a an independent field and if we were to consider
just the first integral in (10), this would be the ordinary Yang
Mills theory (after solving $\delta$ and inserting back into the
action). The second integral represents then the departure from
the usual Yang Mills theory. Here $\omega$ is not an independent
field, but is again defined in terms of the three index potential
\begin{equation}
\omega = \epsilon^{\lambda\mu\nu\rho}
\partial_{[\,\lambda}\, A_{\mu\nu\rho \,]}
\end{equation}
In order to have scale invariance the potential must be quartic,
that is
\begin{eqnarray}
V(\phi)= g \phi^{4}
\end{eqnarray}

In addition to the scale transformation (2) for the gauge field,
 $\delta$, $\phi$ and
$A_{\mu\nu\rho}$ have the scaling transformations
\begin{eqnarray}
&&\delta  \mapsto \lambda ^{-2} \delta \left(\, \lambda\, x\right)\label{cr25}\\
&& \phi \mapsto \lambda ^{-1} \phi\left( {\lambda x}\right)\label{cr25b}\\
&&A_{\mu\nu\rho}\mapsto \lambda ^{-1} A_{\mu\nu\rho}\left(
{\lambda x}\right) \label{cr25c}
\end{eqnarray}

Here we have included a new dynamical degree of freedom, the
scalar field $\phi$. This field, as a consequence of the ssb of
scale invariance will acquire an expectation value and then
confinement behavior will be a consequence of this. However, the
expectation value can become zero under certain conditions, like
in the central regions containing colored particles (this has some
resemblance to the MIT bag model) and also due to the high
temperature dynamics of the scalar field that can make the scalar
field $\phi$ lose its expectation value.

Notice that in order to have a non-trivial dynamics for the scalar
field $\phi$ it is necessary to consider not only a linear term in
$\omega$ but also a quadratic term. It can be easily seen that the
quadratic term in $\omega$ is exactly equivalent to a kinetic term
for the three index field $A_{\mu\nu\rho}$. The consideration of
the linear term in $\omega$ alone would force the scalar field
$\phi$ to be a constant, i.e., non dynamical.

It is in fact the dynamics of $\omega$, or more precisely the
dynamics of $ A_{\mu\nu\rho }$, that is the one responsible for the ssb of
scale invariance. Indeed the variation with respect to these
fields gives rise to the equation
\begin{eqnarray}
 \epsilon^{\lambda\mu\nu\rho} \partial_{\lambda}
 \left[\ \omega -\phi^{2} \right]\ = 0
\end{eqnarray}

This can be integrated very simply, giving rise to the solution
\begin{eqnarray}
\omega = \phi^{2} + M
\end{eqnarray}
The constant $M$ gives rise to ssb of scale invariance now.

The equation obtained from the variation of $\delta$ leads to
\begin{eqnarray}
\delta = \sqrt { - F_{\mu \nu }^{a}\, F^{a\mu \nu } } - 2 \phi^{2}
\end{eqnarray}
and the equation obtained from the variation with respect to the
$\phi$ field is
\begin{eqnarray}
\Box \phi + 4g \phi^{3} -2 \phi \left[ \ \omega - \delta \right] =
0
\end{eqnarray}
Using eqs. (17) and (18) in (19), we obtain
\begin{eqnarray}
\Box \phi + 4g \phi^{3} - 2 \left[ \ 3 \phi^2 -  \sqrt { - F_{\mu
\nu }^{a}\, F^{a\mu \nu } } \right]\phi - 2M\phi = 0
\end{eqnarray}

The equation of motion for the gauge field is now
\begin{equation}
 \nabla _ \mu
\left[ {\left( {\sqrt { - F_{\alpha \beta }^a F^{a\alpha \beta } }
 - \phi^{2}} \right
 )\frac{{F^{a\mu \nu } }}{{\sqrt { - F_{\alpha \beta }^b
F^{b\alpha \beta } } }}} \right] = 0. \label{cr52}
\end{equation}

If $\phi$ is almost a constant (and non vanishing), we have the
infrared confining solution (following the same arguments of the
previous section) satisfying
\begin{equation}
{F^{a\mu \nu } =  \phi^{2}\frac{{F^{a\mu \nu } }}{{\sqrt { -
F_{\alpha \beta }^b F^{b\alpha \beta } } }}}. \label{cr53}
\end{equation}

Following the arguments of the previous section, this means that
the string tension is proportional to $\phi^{2}$.
 In this regime we have that
\begin{equation}
\sqrt { - F_{\alpha \beta }^a F^{a\alpha \beta } }
 = \phi^{2}  \label{cr54}
\end{equation}
Using this in the equation for the scalar field we obtain
\begin{eqnarray}
\Box \phi + 4\left(g -1\right)\phi^{3} - 2M\phi = 0
\end{eqnarray}

We see that when taking into account the effects of the gauge
fields, the quartic self coupling of the scalar field is $g-1$
(notice therefore that the perturbative regime is achieved when
$g$ is close to $1$) and the scalar field acquires now the
effective potential
\begin{eqnarray}
V_{eff} \left(\phi\right)=
 \left(g-1\right)\phi^{4} - M\phi^{2}
\end{eqnarray}
We have thus obtained a consistent effective potential for the
dilaton field $\phi$ after classically integrating out the gauge
fields.

We see then that if $g -1> 0$ and $M>0$, then the scalar field
$\phi$ will acquire an expectation value determined by the minimum
of $V_{eff}$
\begin{eqnarray}
 <\phi>=
 \left(\frac{M}{2g-2}\right)^{\frac{1}{2}}
\end{eqnarray}

At high temperatures\cite{highT}, we will expect the effective
potential for $\phi$ to acquire  a term proportional to
$\left(g-1\right)\phi^{2}T^{2}$ and therefore for high enough
temperature compared to
$\left(\frac{M}{2g-2}\right)^{\frac{1}{2}}$, the scalar field will
lose its expectation value, but then also the confining property
is lost.

At zero temperature if we are close to the source of the
chromoelectic field, we see from equation (20) that the square
root term there will become large and the leading term will be
$\phi$ independent, being indeed a Coulomb piece. This has the
effect of stabilizing  $\phi=0$ as the true vacuum once we
approach the chomoelectric source. As in the case of high T,
losing this expectation value means losing the confining piece
of the gauge fields. A bag type picture is obtained then, where
the perturbative behavior holds very close to the sources and
$<\phi>=
 \left(\frac{M}{2g-2}\right)^{\frac{1}{2}}$ holds far
 enough from the source. This $<\phi>$ induces then, in the
 infrared region, the confining behavior.

 It is interesting to note that models that control the high
 temperature dynamics of strongly interacting particles using a
 dilaton field have been studied \cite{Kalbermann}, although in
 this case, in terms of nuclear physics degrees of freedom.

 Other models which use a scalar to generate a confinement mechanism
 have been studied in \cite{Wilets}. Such models are not constrained
 by a symmetry principle like scale invariance. Although very
 interesting,  such models (in contrast
 to our approach) achieve confinement only as the result of a
 fine tuning: in \cite{Wilets}  a dielectric
 function $\kappa$ is introduced which is a function of the
 scalar field and also a  potential $U$ for the scalar field is
 included in the action. For confinement to be realized, there
 must be a value of the scalar field that
 satisfies both that $\kappa= 0$ and at the same time, the value
 of the scalar field that achieves this is located at a minimum of
 $U$ and at an extremum of $\kappa$. This of course can be achieved only
 through a fine tuning.

\section{Concluding Remarks and Discussion}
We have seen that in the model presented here confinement becomes
associated with the non-vanishing expectation value of the field
$\phi$. The expectation value of $\phi$ is proportional to the
string tension between static charges. The introduction of the
scalar field $\phi$ allows us to study the dynamics that can be
responsible for the confinement/deconfinement phase transition,
for example at finite temperature. Also when we get close to the
chromoelectric sources, we expect the expectation value of the
scalar field $\phi$ to become zero.

\vskip.8in

\centerline{{\bf Acknowledgments}} I wish to thank E. Spallucci
and P. Gaete for discussions.

\vskip1.6in


\vskip.3in

\begin{thebibliography}{99}

\bibitem{free}
H.D. Politzer, Phys. Rev. Lett. {\bf 30}, 1346 (1973); Phys.
Rept{\bf 14} 129 (1974); D.J. Gross and F. Wilczek, Phys. Rev.
Lett. {\bf 30}, 1343 (1973); Phys. Rev. {\bf D8}: 3633 (1973);
Phys. Rev. {\bf D9}: 980 (1974).

\bibitem{deconf}
See for example G. Baym, "Deconfined Phases of Strongly
Interacting Matter" in Quark Matter Formation and heavy Ion
Collisions, M. Jacob and H. Satz, World Scientific (1982);
 For an up to date review see J.P. Blaizot,
nucl-th/0611104.

\bibitem{lattice}
K. G. Wilson, Phys. Rev. {\bf D10}, 2445 (1974). The literature in
this subject is huge and the uses of lattice techniques very
extended, for example, for an interesting recent study of the
phases of  gauge theories in the context of these lattice
techniques see M. Grady, Contributed to 24th International
Symposium on Lattice Field Theory (Lattice 2006), Tucson, Arizona,
23-28 Jul 2006. e-Print Archive: hep-lat/0610042.

\bibitem{Cornell}
E. Eichten, K. Gottfried, T. Kinoshita, K. D. Lane,
 and T. M. Yan, Phys. Rev. {\bf D17}, 3090 (1978).

\bibitem{MIT}
A. Chodos, R.L. Jaffe, K. Johnson, C.B. Thorn and V.F. Weisskopf,
Phys. Rev. {\bf D9}, 3471 (1974).


\bibitem{Guendelman}
E.I. Guendelman, Int. J. Mod. Phys.{\bf A19} 3255 (2004).

\bibitem{pert-conf}
 G. 't Hooft, Nucl. Phys. Proc. Suppl. {\bf
 121}, 333 (2003).

\bibitem{Guendelman-Gaete}
P. Gaete and E. I. Guendelman, Phys. Lett. {\bf B640}, 201 (2006);
P. Gaete, E.I. Guendelman and E. Spallucci, hep-th/0702067, to
appear in Phys. Lett. {\bf B}.


\bibitem{interplay}
P. Gaete and E. I. Guendelman, Phys. Lett.{\bf B593} 151 (2004).

\bibitem{highT}
L.Dolan and R. Jackiw, Phys. Rev. {\bf D9} 3320 (1974); S.
Weinberg, Phys. Rev. {\bf D9} 3357 (1974).

\bibitem{Kalbermann}
G. Kalbermann, J.M. Eisenberg and B. Svetitsky, Nucl. Phys. A600:
436 (1996).

\bibitem{Wilets} See for example W. Koepf, L. Wilets and S. Pepin, Phys. Rev. {\bf
C50}: 614 (1994).

\end{thebibliography}
\end{document}